\documentclass[aps,prb,reprint,longbibliography,nofootinbib,floatfix]{revtex4-2}

\usepackage{amsmath,amssymb,mathtools}
\usepackage{graphicx}
\usepackage{array}
\usepackage{microtype}
\usepackage[dvipsnames]{xcolor}
\usepackage[colorlinks=true,linkcolor=MidnightBlue,citecolor=MidnightBlue,urlcolor=MidnightBlue]{hyperref}

\graphicspath{{figures/}{figures/schematics/}{figures/data/grace/}}

\newcommand{\Tr}{\operatorname{Tr}}
\newcommand{\NH}{N_{\mathrm H}}

\begin{document}

\title{Hartree--Fock Density-Matrix Renormalization Group for Very Long Chains}

\author{Steven R. White}
\affiliation{Department of Physics and Astronomy, University of California, Irvine, Irvine, CA 92697, USA}

\date{September 14, 2026}

\begin{abstract}
We formulate Hartree--Fock as a density-matrix renormalization group
(DMRG) sweep of a single Slater determinant, assuming a density--density
electron--electron interaction. We compress the occupied state and
long-range interaction separately, reducing each sweep step to a small
local Fock problem. Exchange is retained throughout. Uniform
antiferromagnetic hydrogen chains provide an almost ideal setting for
this approach, while retaining the Coulomb interaction and its $1/r$
tail. We initialize and converge a chain of 100,000 electrons in
1.8 million spatial basis functions in about two hours on a 64-GB Mac mini,
using less than a third of its memory. The work per sweep is nearly
proportional to chain length, and the number of sweeps needed for
convergence changes little from 1000 to 100,000 atoms. DMRG's compression
of states and environments thus makes large HF calculations
with long-range Hartree and exchange practical on an ordinary desktop
computer.

\end{abstract}

\maketitle

\section{Introduction and background}
\label{sec:introduction}

Hartree--Fock (HF) is one of the simplest approximations to the electronic
many-body problem, but solving it for a very large system can still be
expensive. In the usual self-consistent-field (SCF) procedure, the occupied
orbitals determine a Fock matrix, whose lowest eigenvectors give the next
set of orbitals. The cost of dense diagonalization grows as the cube of the basis size,
and even the occupied-orbital coefficient matrix grows quadratically when
the number of electrons increases with system size. The long-range Coulomb
and exchange fields add another source of cost. These difficulties matter
both for HF itself and for related orbital methods, including Kohn--Sham
density functional theory \cite{KohnSham1965}.

Considerable work has gone into removing this unfavorable scaling
\cite{Goedecker1999}. The
underlying physical idea is locality: in an insulator, distant changes
usually have a small effect on the local occupied state
\cite{ProdanKohn2005}. Density-matrix minimization
\cite{LiNunesVanderbilt1993} and purification \cite{Niklasson2002}
avoid global diagonalization, using sparse matrices to obtain linear
scaling when the density matrix is sufficiently local. Orbital-minimization
methods vary localized orbitals directly \cite{MauriGalliCar1993},
while divide-and-conquer
methods instead determine the density from overlapping subsystems
\cite{Yang1991}. Long-range electrostatics can be treated by fast multipole
methods \cite{WhiteCFMM1994}, while screening, incremental construction,
and local density fitting reduce the cost of exchange
\cite{Schwegler1997,KopplWerner2016}. Combinations of these ideas have
led to linear-scaling molecular SCF implementations
\cite{Salek2007}. Both the construction of the fields and the update of
the occupied state must be inexpensive for a complete HF calculation
to scale well.

Here we take a different route, following the finite-system density-matrix
renormalization group (DMRG) \cite{White1992,White1993}. In a DMRG sweep,
a small region is optimized between compressed descriptions of the
system to its left and right. With fixed state and matrix-product-operator
(MPO) dimensions, each step has fixed cost and a sweep is linear in the
length \cite{Schollwock2011}. The state dimension is particularly
favorable for gapped one-dimensional systems. Long-range interactions
can also be inexpensive if their couplings across a cut have low rank.
The natural question is whether the same organization can make HF
inexpensive, with the Slater determinant replacing the many-body
matrix product state (MPS) \cite{OstlundRommer1995}.

The connection between DMRG and one-particle problems is longstanding.
A single-particle tight-binding problem helped expose the importance of
including the environment when choosing a block basis
\cite{WhiteNoack1992}. For a filled set of free-fermion orbitals, the
many-body reduced density matrix can be obtained from the much smaller
one-particle correlation matrix restricted to a block
\cite{ChungPeschel2001,Peschel2003,CheongHenley2004}. Its spectrum
identifies occupied modes shared by the two sides of a cut
\cite{BoteroReznik2004}. Most modes of a local state are almost filled
or almost empty within the block, so only a few need to be retained.
This is the one-particle version of the density-matrix truncation at
the heart of DMRG.

Fishman and White gave an efficient algorithm for the compression of
correlation matrices using local orbital rotations, and gave a DMRG-like
algorithm for noninteracting many-fermion systems with local hoppings
\cite{FishmanWhite2015}. Schuch and Bauer
developed a broader set of Gaussian fermionic MPS algorithms, demonstrating
free-fermion calculations with up to a million sites
\cite{SchuchBauer2019}. Meiburg and Bauer then used a Gaussian-MPS solver
inside a generalized-HF self-consistency loop, obtaining approximately
linear scaling for quasi-one-dimensional Hubbard systems
\cite{MeiburgBauer2022}. In that algorithm, several sweeps solve a fixed
quadratic Fock problem before the field is rebuilt. Their treatment
also makes clear the difficulty posed by unscreened Coulomb interactions:
extracting all the density-matrix entries needed for a dense Fock matrix
would lose the advantage of the compressed state.

HF-DMRG avoids forming this dense Fock matrix by combining the compressed
determinant with compressed long-range interaction fields. It follows the
finite-system DMRG algorithm closely, with groups of adjacent basis
functions playing the role of single sites. Each step optimizes two such
blocks between compressed left and right environments, then advances the
center and recalls the next environment from memory. At the end of the
chain, the environments for the return sweep are ready, just as in DMRG.
We rebuild the small Fock problem at each position, so orbital optimization
and self-consistency proceed through the same sweeps.
The calculation uses direct sums of one-particle spaces rather than tensor
products of many-body spaces. Its compression is based on low rank across
a cut, rather than a distance beyond which every density-matrix element
is set to zero. This gives a different organization of linear-scaling HF;
our aim here is to establish the accuracy and scaling of that organization
for long chains.

We assume a two-index density--density interaction. This form is natural
on a real-space grid and is also an accurate approximation in a gausslet
basis \cite{White2017Gausslets,WhiteStoudenmire2019}. It retains the
long-range Coulomb interaction and HF exchange, while greatly simplifying
the operator information carried by a block. We demonstrate the method
on hydrogen chains in a fine longitudinal basis with one fixed transverse
function at each longitudinal site. The uniform antiferromagnetic chain is close to ideal
for the algorithm: its occupied state is local, its interaction factors
can be reused along the chain, and a simple atomic product is a good
starting state. Nevertheless, every electron interacts with the whole
chain through the realistic Coulomb tail, and we can retain high numerical
precision.

For this family, a chain of 100,000 electrons and 1,800,003 spatial
functions can be initialized and converged in about two hours on a 64-GB
Mac mini, using about 21 GB. The number of sweeps needed for convergence
changes very little between 1000 and 100,000 atoms. State compression
also reduces the orbital-pair space needed for exchange, leaving a typical
local Fock matrix with only about 45 rows throughout this range of
lengths. Adding atoms mainly adds sweep steps, not larger local problems.

We first specify the Hamiltonian and chain basis in Sec.~\ref{sec:model}.
Sections~\ref{sec:methods-sweep} and \ref{sec:methods-optimizer} follow the
DMRG sweep, explaining the three compressions and the local HF update
that make it inexpensive. We then examine their accuracy and the
convergence and scaling of the long chains in
Secs.~\ref{sec:accuracy} and \ref{sec:convergence}, before discussing
the role of dimensionality and possible extensions.

\section{Hamiltonian and chain representation}
\label{sec:model}

We use $N$ real orthonormal basis functions ordered along a chain. Site
indices $i,j,k$ label individual basis functions; block indices $a,b,c$
label groups of these functions. We use $p,q,r,s$ for orbitals within
a retained or occupied space, and $\sigma,\tau\in\{\uparrow,\downarrow\}$
for spin-up and spin-down. Matrices are
bold, vectors carry arrows, and scalar entries are plain: for example,
$V_{ij}$ is one entry of the matrix $\mathbf V$. Calligraphic letters
denote one-particle spaces. Let
$c^\dagger_{i\sigma}$ create an electron of spin $\sigma$ in function $i$.
The Hamiltonian is
\begin{align}
 \hat H={}&E_{\rm NN}+\sum_{ij\sigma}h_{ij}
 c^\dagger_{i\sigma}c_{j\sigma} \nonumber\\
 &+\frac12\sum_{ij\sigma\tau}V_{ij}
 c^\dagger_{i\sigma}c^\dagger_{j\tau}c_{j\tau}c_{i\sigma}.
 \label{eq:hamiltonian}
\end{align}
Here the $N\times N$ matrix $\mathbf h$ contains kinetic energy and nuclear
attraction, and $E_{\rm NN}$ is the nuclear repulsion. The density--density
approximation reduces the usual four-index interaction to the symmetric
$N\times N$ matrix $\mathbf V$, often called a diagonal interaction because
it couples density operators. The matrix itself is dense and long-ranged.
The normal-ordered form of Eq.~\eqref{eq:hamiltonian} also specifies the
$i=j$ terms without self-interaction.

The orthonormal occupied orbitals form the columns of the $N\times n_{\mathrm{occ},\sigma}$
matrix $\mathbf C^\sigma$, where $n_{\mathrm{occ},\sigma}$ is the number
of electrons of spin $\sigma$. The one-particle density matrix is the
$N\times N$ projector $\mathbf D^\sigma=\mathbf C^\sigma(\mathbf C^\sigma)^T$.
Unrestricted
HF (UHF) allows independent spin projectors; restricted HF (RHF) sets
them equal. With total density
$\rho_i=D^\uparrow_{ii}+D^\downarrow_{ii}$ forming the length-$N$ vector
$\vec\rho$, the energy and $N\times N$ Fock matrices are
\begin{align}
 E={}&E_{\rm NN}+\sum_\sigma\Tr(\mathbf h\mathbf D^\sigma)
       +\frac12 \vec\rho^{\,T}\mathbf V\vec\rho \nonumber\\
    &-\frac12\sum_{ij\sigma}V_{ij}(D^\sigma_{ij})^2,
 \label{eq:hf-energy}\\
 \mathbf F^\sigma={}&\mathbf h+\operatorname{diag}(\mathbf V\vec\rho)
                         -\mathbf V\circ\mathbf D^\sigma,
 \label{eq:fock}
\end{align}
where $\circ$ denotes elementwise multiplication. The last term in each
equation is exchange. Thus the two-index interaction simplifies exchange
as well as Hartree: both use the same $\mathbf V$, but exchange also contains the
off-diagonal same-spin density matrix. We keep both terms at all distances.

Our main example is an open-ended hydrogen chain with spacing $R=3.6$ bohr and
one electron per atom. The spacing is large compared to the H$_2$
equilibrium bond length, making the system resemble an antiferromagnetic
spin chain. Following the sliced-basis approach of Sawaya and White
\cite{sawayawhite}, we use basis functions of the form
\begin{equation}
 \psi_i(x,y,z)=g_i(z)\phi_i(x,y),
 \label{eq:chain-basis}
\end{equation}
where $g_i$ is a G10 gausslet centered at $z_i$, with centers spaced by
$R/18$ (0.2 bohr at $R=3.6$), and $\phi_i$ is one normalized, cylindrically symmetric
transverse function. To construct $\phi_i$, we restrict each atom-centered
STO-6G hydrogen $1s$ function to the plane $z=z_i$, sum the outer products
of these restricted functions, and take the eigenfunction with the largest
eigenvalue. This auxiliary density depends only on the atomic basis and
geometry. For the uniform chains we construct the transverse functions
from a periodic array of atoms and repeat their pattern in each atomic
interval, keeping them fixed during HF optimization of the finite,
open-ended chain. The nuclei lie on gausslet centers, and the grid extends
2 bohr beyond each terminal nucleus, rounded outward to a gausslet center.
We keep 18 functions per atomic interval when varying $R$. At $R=3.6$ bohr,
the chosen end padding gives
\begin{equation}
 N=18\NH+3,
 \label{eq:basis-size}
\end{equation}
where $N$ is the spatial basis size and $\NH$ the number of atoms.

We evaluate the kinetic and nuclear-attraction matrix elements in this
basis. Products of well-separated gausslets decay rapidly, allowing
distant off-diagonal entries of $\mathbf h$ to be omitted using error
bounds. This gives a symmetric banded matrix, with a separate one-body
accuracy tolerance of $10^{-8}$ Ha for the long chains. The nuclear
attraction includes all nuclei; the band limits the separation between
the two basis functions in a matrix element.

For the interaction we use the integral diagonal approximation along
the chain \cite{White2017Gausslets,StoudenmireWhite2017,WhiteStoudenmire2019}.
Writing $I_i=\int g_i(z)\,dz$ and $\vec r=(x,y,z)$, this gives
\begin{equation}
 \begin{aligned}
 V_{ij}={}&\frac{1}{I_i I_j}
   \int\frac{d^3r\,d^3r'}{|\vec r-\vec r'|}\\
 &\quad\times g_i(z)g_j(z')|\phi_i(x,y)|^2|\phi_j(x',y')|^2.
 \end{aligned}
 \label{eq:integral-diagonal-interaction}
\end{equation}
The longitudinal averaging uses the weights $g_i/I_i$, while the
transverse factors are normalized densities.
The resulting interaction has the physical $1/r$ tail, with its
short-distance behavior set by the basis functions. The fixed transverse
space and diagonal approximation define the model; the numerical
precision discussed below is within this model, not a complete-basis
accuracy for a three-dimensional molecule.

The stretched chain favors an antiferromagnetic UHF state. Its locality
makes it a good first test of how far HF-DMRG can scale. We also compare
RHF and shorter chains at smaller spacing, where more orbitals cross each
cut. All large calculations have equal spin populations, $\NH/2$ in
each sector.

\section{Sweeping and compression}
\label{sec:methods-sweep}

We group the ordered basis functions into $n_b$ blocks $B_a$ of size
$d_a=|B_a|$. Their one-particle spans are denoted by $\mathcal B_a$.
Each block plays the role of one DMRG site. It can contain an atom's
basis functions or simply a convenient interval of the original basis.
At position $a$, we solve the local problem in
\begin{equation}
 \mathcal H_{\sigma,a}=
 \mathcal L_{\sigma,a-1}\oplus\mathcal B_a\oplus\mathcal B_{a+1}
 \oplus\mathcal R_{\sigma,a+2},
 \label{eq:methods-center-space}
\end{equation}
where $\mathcal L$ and $\mathcal R$ are the retained left and right
one-particle spaces. Their dimensions are $m_{\sigma,L}$ and
$m_{\sigma,R}$. Since these spaces add rather than multiply, the Fock
matrix has only $m_{\sigma,L}+d_a+d_{a+1}+m_{\sigma,R}$ rows and columns.
Here and below these retained dimensions refer to the current position.
For the $R=3.6$ hydrogen chains,
with 18-function blocks and four or five retained modes on each side, a
typical bulk problem has 44--46 rows, independent of the chain length.

\begin{figure}[t]
\centering
\includegraphics[width=\columnwidth]{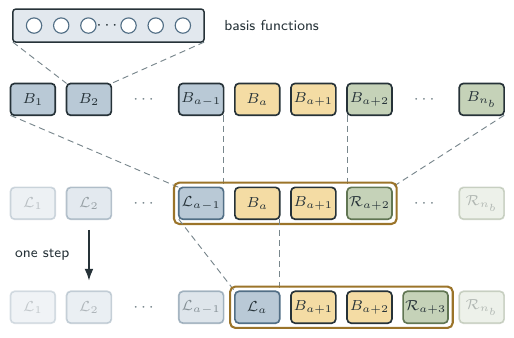}
\caption{\label{fig:sweep}An HF-DMRG sweep.
The ordered basis is partitioned into blocks $B_a$. At a given sweep position,
the blocks preceding and following the two-block center are compressed into
the stored environments $\mathcal L_{a-1}$ and $\mathcal R_{a+2}$, with
spin labels suppressed. Advancing the sweep absorbs
$B_a$ into $\mathcal L_{a-1}$ to form $\mathcal L_a$ and retains $B_{a+1}$ in the center; the
already stored $\mathcal R_{a+3}$ is recalled for the next local problem. Earlier stored
environments are drawn faintly. The upper inset shows how one block groups
an interval of the ordered basis.}
\end{figure}

Figure~\ref{fig:sweep} shows a step of the sweep. After improving the
determinant in this space, we absorb $B_a$ into the left environment,
keep $B_{a+1}$ in the center, and recall the next right environment from
storage. The new left environment replaces the consumed right environment
at that position, so a single collection holds the environments for both
directions. At the end of the chain, we factor the last optimized center
with the partition appropriate to the return sweep. The center remains
available for optimization, and the environments it will meet were built
on the outward sweep. A one-block version is also possible, as in one-site
DMRG; we use two blocks so that the retained space can grow and adjust
as the state changes.

\subsection{Three types of compression}
\label{sec:three-compressions}

For the algorithm to be nearly linear in cost with the basis size, the
work at each position must be nearly constant.
This requires compressing both the occupied state
and the interaction information carried by the left and right blocks.
Three types of compression can be used, acting on different objects.

\emph{(1) One-particle compression} retains the orbital modes shared
across a block boundary, separating them from modes that are filled
or empty within the block. This is the one-particle analogue of DMRG
state truncation. It determines the left and right orbital spaces
in Fig.~\ref{fig:sweep}. The key to this compression is rotating the bases
on the left or right to make them as occupied or unoccupied as possible.
Once a mode is filled or empty to the specified accuracy, we call it
completed. It drops out of the local optimization, with its energy and
fields incorporated into the environment, until the sweep returns to
that position.

\emph{(2) MPO compression} retains the important combinations of
interaction terms coupling densities across a cut. As in ordinary
DMRG, for high efficiency a compact representation of the state must be accompanied by
a compact representation of the Hamiltonian.

\emph{(3) Density-space compression} acts on products of retained
orbital functions, which enter the projected Hartree and exchange
integrals. It selects combinations needed to represent the state
and its operator responses, much as DMRG can target several states.
We also call this \emph{pair-response compression}.

In our long hydrogen chains, the first two compressions work so well
that the remaining orbital-product space can be retained in full.
The third compression can be useful when that space is larger. We
describe the three in this order below.

\subsection{One-particle compression}

We begin with the left and right orbital spaces in Fig.~\ref{fig:sweep}.
Which combinations of orbitals within a block must remain available
for optimization with the rest of the chain? To answer this, split
one spin projector at a cut:
\begin{equation}
 \mathbf D^\sigma=
 \begin{pmatrix}\mathbf D^\sigma_{LL}&\mathbf D^\sigma_{LR}\\
 (\mathbf D^\sigma_{LR})^T&\mathbf D^\sigma_{RR}\end{pmatrix},
 \qquad (\mathbf D^\sigma)^2=\mathbf D^\sigma.
 \label{eq:methods-projector-cut}
\end{equation}
Here $L$ and $R$ are sets of physical sites on the two sides of the cut,
containing $N_L$ and $N_R$ sites, with $N_L+N_R=N$.
The diagonal restrictions $\mathbf D^\sigma_{LL}$ and $\mathbf D^\sigma_{RR}$
have sizes $N_L\times N_L$ and $N_R\times N_R$; the cross-cut matrix
$\mathbf D^\sigma_{LR}$ has size $N_L\times N_R$.
The eigenvalues of the left restriction are the mean occupations
$\langle n\rangle$ of its modes, lying between zero and one.
An eigenvalue one describes an occupied mode entirely on the left, and
zero describes an empty mode. Intermediate eigenvalues describe occupied
modes shared across the cut. These are the modes the compressed block
must pass to the next local problem. For a free-fermion state, their
occupations determine the many-body Schmidt weights
\cite{ChungPeschel2001,Peschel2003,CheongHenley2004,BoteroReznik2004}.
If $m$ independent occupied modes straddle the cut, their occupation
patterns can give $2^m$ many-body Schmidt weights. Keeping the $m$
one-particle modes avoids enumerating these patterns, making the Gaussian
representation much less expensive than a general MPS at the same
entanglement \cite{SchuchBauer2019}.

The singular-value decomposition (SVD) of the off-diagonal block contains
closely related information. Idempotency gives
\begin{equation}
 \begin{aligned}
 \mathbf D^\sigma_{LR}(\mathbf D^\sigma_{LR})^T
   &=\mathbf D^\sigma_{LL}-(\mathbf D^\sigma_{LL})^2,\\
 \lambda_p^2&=\langle n\rangle_p(1-\langle n\rangle_p),
 \end{aligned}
 \label{eq:methods-cross-projector}
\end{equation}
where $\lambda_p$ is the singular value of $\mathbf D^\sigma_{LR}$ corresponding
to mode occupation $\langle n\rangle_p$. Small singular
values therefore identify modes that communicate little across the cut.
The diagonal block is still needed: $\langle n\rangle$ and
$1-\langle n\rangle$ have the same singular value,
so the SVD alone cannot distinguish filled from empty modes. In UHF we
make this decomposition separately for each spin. The sum
$\mathbf D^\uparrow+\mathbf D^\downarrow$ is not generally a projector.

Local rotations that isolate filled and empty modes give a practical
way to construct the compressed state \cite{FishmanWhite2015}.
As in DMRG, we retain the shared modes subject to a
maximum dimension and a bound on the combined discarded weight at each cut.
For one spin, let $w_p$ be the probability lost by making mode $p$
filled or empty. The combined weight is
$w_{\rm disc}=1-\prod_p(1-w_p)$, with the product over all such modes.
For small losses this is approximately $\sum_p w_p$, the sum of their
distances from occupations zero or one.
The cutoff applies to this combined weight, rather than to each occupation
separately. When the dimension limit does not bind, we retain the smallest
active space meeting the weight bound. Clusters of nearly degenerate singular values
are never split, making the retained space independent of the choice
of singular vectors within a cluster.

Discarded weight controls representation of the state, not the energy
directly. Changes in the associated projector scale as its square root.
Near a stationary determinant, the leading energy change is quadratic
in the orbital change and is therefore expected to scale with discarded
weight. We measure the energy error separately.

Efficiency comes from removing the filled and empty modes from subsequent
local diagonalizations. As a mode becomes fully occupied within the
growing block, we incorporate its one-body energy and its Hartree and
exchange fields into that block. These contributions remain valid while
the completed orbitals are held fixed, even as we optimize the active
orbitals elsewhere. We also store the orbital transformation as a
reversible link, so that the mode can reenter the center on the return
sweep and its fields can be updated as it is optimized.

A stored block contains this state link, the projected one-body
Hamiltonian, and the interaction and completed-energy terms described
below. The occupied state at the moving center is a small density
matrix. Together the center and links specify the full determinant,
without storing an $N\times n_{\mathrm{occ},\sigma}$ coefficient matrix
for either spin. The local factorizations use the occupied coordinates
present in the center. Orbitals completed elsewhere contribute through
their counts and stored fields, keeping the local work independent of
the total occupation.

We start from local products, as is often done in DMRG. For UHF, we
partition the basis at the midpoints between nuclei, restrict the one-body
matrix $\mathbf h$ to each atomic interval, and occupy its lowest eigenvector
with alternating spins. For RHF, we join neighboring intervals into
disjoint dimers and doubly occupy the lowest eigenvector of each
restricted $\mathbf h$. These matrices retain the nuclear attraction
of the whole chain; initialization requires only local one-body
diagonalizations, with no preliminary HF optimization. The occupied
orbitals have disjoint supports, so their block representations can be
built directly. The sweep then adjusts them without constructing global
coefficients at initialization or between sweeps.

\subsection{MPO compression of the interaction}
\label{sec:methods-interaction}

A local state is not enough to make the calculation linear: each density
still interacts with every other density through $V_{ij}$. DMRG handles
this problem by compressing the Hamiltonian as well as the state.
For example, an exponential interaction in separation propagates from
one unit cell to the next by a fixed scalar decay factor. Approximating
the Coulomb tail by a sum of exponentials therefore needs only one
channel per term \cite{CrosswhiteDohertyVidal2008}. Successive SVDs find the important
channels for more general interactions, as we now describe.

\begin{table}[htbp]
\caption{\label{tab:mpo-dimensions}Matrix sizes at the cut after block $B_a$.
Here $d_a$ is the number of sites in the block, $N_L$ and $N_R$ count
all sites on each side of the cut, and $\chi_a$ is the retained channel
rank. Only the four local matrices in the lower group are stored in
the finished MPO.}
\begin{ruledtabular}
\begin{tabular}{ll}
Matrix & Rows $\times$ columns \\
$\mathbf L_a$ & $N_L\times\chi_a$ \\
$\boldsymbol\Sigma_a$ & $\chi_a\times\chi_a$ \\
$\mathbf W_a$ & $N_R\times\chi_a$ \\
$\mathbf X_a$ & $\chi_a\times N_R$ \\
$\mathbf U_a$ & $(d_a+\chi_{a-1})\times\chi_a$ \\
\hline
$\mathbf V_{aa}$ & $d_a\times d_a$ \\
$\mathbf S_a^T$ (source) & $d_a\times\chi_a$ \\
$\mathbf T_a$ (transfer) & $\chi_{a-1}\times\chi_a$ \\
$\bar{\mathbf S}_a$ (sink) & $\chi_{a-1}\times d_a$
\end{tabular}
\end{ruledtabular}
\end{table}

\begin{figure}[!t]
\includegraphics[width=\columnwidth]{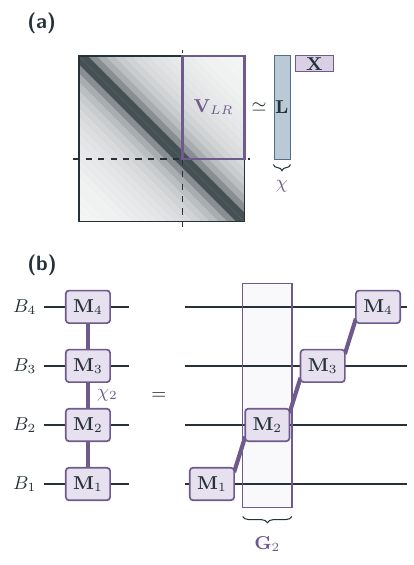}
\caption{\label{fig:mpo}Compression of the long-range interaction.
(a) Truncating the singular-value decomposition of the cross-cut block gives
$\mathbf V_{LR}\simeq\mathbf L_a\mathbf X_a$ with $\chi_a$ retained
channels; the cut index $a$ is suppressed in the drawing.
(b) Two drawings of the same direct-sum network. The left diagram uses the
usual MPO layout; the right shows the corresponding product of matrices
over the whole space. Thin wires carry
physical-block coordinates and thick links carry retained channels; parallel
space dimensions add rather than multiply. The outlined slice is $\mathbf G_2$:
the local matrix $\mathbf M_2$ together with identity actions on the other three
blocks. Reading left to right in a row-vector convention gives
$\mathbf V^\triangle\simeq\mathbf G_1\mathbf G_2\mathbf G_3\mathbf G_4$,
where $\mathbf V^\triangle$ contains the full diagonal
blocks and upper off-diagonal blocks; symmetry supplies the lower blocks.
The local matrices $\mathbf M_a$ of Eq.~\eqref{eq:methods-local-dsn}
contain local, source, sink, and transfer blocks.}
\end{figure}

The scalar interactions $V_{ij}$ between sites in blocks $B_a$ and $B_b$
form the matrix
\begin{equation}
 \mathbf V_{ab}=[V_{ij}]_{i\in B_a,\,j\in B_b},
 \qquad \mathbf V_{ab}\in\mathbb R^{d_a\times d_b}.
 \label{eq:methods-block-interaction}
\end{equation}

Across the cut after $B_a$, all left--right interactions lie in the
$N_L\times N_R$ matrix $\mathbf V_{LR}$, with
$N_L=\sum_{c\le a}d_c$ and $N_R=\sum_{c>a}d_c$.
Its truncated SVD is
\begin{equation}
 \mathbf V_{LR}\simeq\mathbf L_a\boldsymbol\Sigma_a\mathbf W_a^T
       \equiv\mathbf L_a\mathbf X_a.
 \label{eq:methods-cross-interaction}
\end{equation}
Here $\chi_a$ is the number of channels retained at this cut,
$\boldsymbol\Sigma_a$ is the $\chi_a\times\chi_a$ diagonal
matrix of retained singular values, and $\mathbf W_a$ contains the
$\chi_a$ right singular vectors, each of length $N_R$.
The long factor $\mathbf L_a$ contains the corresponding left singular
vectors: it has one row for every site left of the cut and $\chi_a$
orthonormal columns, so $\mathbf L_a^T\mathbf L_a=\mathbf I_{\chi_a}$.
Each singular-vector pair couples one density pattern on the left to one on
the right. A channel therefore represents a weighted sum of site-density
operators, whose expectation value supplies one component of the Hartree
field passed to the other side. Only these $\chi_a$ channels need to pass
between the blocks, playing the role of renormalized operators in DMRG.
This is the same low-rank structure used to make compact long-range
MPOs \cite{LinTong2021}. It is shown schematically in
Fig.~\ref{fig:mpo}(a).

We construct the MPO by successive SVDs, obtaining small matrices that create,
propagate, and terminate the interaction channels, as derived in
Appendix~\ref{app:sequential-svd}. At step $a$, a smaller SVD combines
the interaction rows of $B_a$ with the $\chi_{a-1}$ incoming channels.
Its left singular-vector matrix $\mathbf U_a$ has
$d_a+\chi_{a-1}$ rows and $\chi_a$ columns.
The two SVDs have the same nonzero singular values and right singular
vectors before truncation. After earlier truncations, the smaller SVD
applies to the interaction carried forward by the preceding steps.
Writing it as $\mathbf U_a\boldsymbol\Sigma_a\mathbf W_a^T$, we define
\begin{equation}
 \mathbf U_a=\begin{pmatrix}\mathbf S_a^T\\\mathbf T_a\end{pmatrix},
 \qquad \mathbf X_a=\boldsymbol\Sigma_a\mathbf W_a^T.
 \label{eq:methods-channel-definitions}
\end{equation}
The first $d_a$ rows of $\mathbf U_a$ give the source matrix
$\mathbf S_a^T$; the remaining rows give the transfer matrix
$\mathbf T_a$. The sink matrix $\bar{\mathbf S}_a$ consists of the first
$d_a$ columns of the preceding right factor $\mathbf X_{a-1}$, coupling
incoming channels to $B_a$; the bar labels the sink.
Table~\ref{tab:mpo-dimensions} lists the dimensions.
For blocks $a<b$, the $d_a\times d_b$ interaction matrix is
\begin{equation}
 \mathbf V_{ab}\simeq
 \mathbf S_a^T\mathbf T_{a+1}\cdots\mathbf T_{b-1}\bar{\mathbf S}_b.
 \label{eq:methods-source-transfer-sink}
\end{equation}
For neighboring blocks the transfer product is absent.
The diagonal block $\mathbf V_{aa}$ is kept separately.

Figure~\ref{fig:mpo} gives a matrix-network picture of the construction.
A local matrix
\begin{equation}
 \mathbf M_a=\begin{pmatrix}
 \mathbf V_{aa}&\mathbf S_a^T\\\bar{\mathbf S}_a&\mathbf T_a
 \end{pmatrix}
 \label{eq:methods-local-dsn}
\end{equation}
has size $(d_a+\chi_{a-1})\times(d_a+\chi_a)$. In the row-vector
convention of the diagram, it connects the physical block and incoming
channels to the physical block and outgoing channels.
Parallel wire dimensions add, rather than
multiply as in a tensor network. The two drawings in the figure are
therefore a \emph{direct-sum} network: the vertical arrangement emphasizes
the MPO-like propagation, while the diagonal arrangement shows ordinary
matrix multiplication. Extending $\mathbf M_a$ by identity actions on the
untouched physical blocks gives a matrix $\mathbf G_a$ of size
$(N+\chi_{a-1})\times(N+\chi_a)$. With zero channel dimensions
at the two endpoints, $\mathbf G_1\cdots\mathbf G_{n_b}$ gives the
$N\times N$ block upper triangle of $\mathbf V$, including the diagonal
blocks. Symmetry supplies the lower off-diagonal blocks. These extended
matrices explain the diagram; only the small $\mathbf M_a$ is stored.
The SVD makes $\mathbf U_a$ isometric, i.e., its columns are orthonormal.

For a uniform chain, we can reuse the factors within a unit cell.
Our interaction combines the exponential-sum Coulomb tail with a
block-Hankel representation of the remaining short-range structure.
For the correction, a block-Hankel matrix groups residual interaction
blocks by the sum of their left and right cell distances from a cut;
its SVD supplies a compact set of additional channels.
We fit this correction over a finite range of cell separations and
use the resulting transfer matrix at larger separations as well.
For fixed ranks, the stored unit-cell operator is independent of length.
Extending the Coulomb tail over a longer range slowly increases the
number of exponentials. The interaction thus remains compact even
though no distance cutoff is imposed. We specify the MPO accuracy by a
target absolute error $\epsilon_V$ in the individual interaction matrix
elements $V_{ij}$, separately from the state truncation. We check the
finished interaction against uncompressed values at all inequivalent
site pairs and separations in the finite chain. The fit range is a
separate choice: tightening $\epsilon_V$ changes the compression within
that construction but does not extend the fitted range. Its effect on
the energy is assessed separately in Sec.~\ref{sec:accuracy}.

\subsection{Density-space compression and exchange}
\label{sec:methods-pairs}

The compact interaction must now be projected into the retained
orbital spaces. In particular, the exchange term $-\mathbf V\circ\mathbf D^\sigma$
is simple in the physical basis, but its block representation involves
products of orbital functions.
Let $\varphi_p(i)$ be the value at physical site $i$ of retained real
orbital $p$, with $p=1,\ldots,m_\sigma$ for one spin. We suppress its
spin label here, and the side and position labels on $m_\sigma$.
When the spin is also understood, we write $m$ and $K$ for the state and
pair dimensions. The pointwise products $\varphi_p(i)\varphi_q(i)$ are the
transition densities needed for the interaction. Since pairs $(p,q)$
and $(q,p)$ give the same product, there are
\begin{equation}
 K_\sigma=m_\sigma(m_\sigma+1)/2
 \label{eq:methods-pair-dimension}
\end{equation}
distinct pairs. For example, the interaction internal to a left block
is a $K_\sigma\times K_\sigma$ matrix $\boldsymbol\Gamma^\sigma$ with
scalar entries
\begin{equation}
 \Gamma^\sigma_{pq,rs}=\sum_{i,j\in L}
 \varphi_p(i)\varphi_q(i)V_{ij}\varphi_r(j)\varphi_s(j),
 \label{eq:methods-pair-matrix}
\end{equation}
where each row and column is labeled by an unordered orbital pair,
chosen as $p\le q$ and $r\le s$.
The same transition densities couple to the outgoing MPO channels.
Retaining all the symmetric pairs is exact within the chosen block
basis.

Keeping the spin spaces independent is important. A direction that
is fully occupied for one spin can still be shared across the cut for
the other.
Compressing the common orbital span would force both spins to carry
directions needed by only one, and the pair dimension amplifies this
extra cost. We instead keep separate spin state links, one-body
operators, and exchange matrices. Hartree couples the two spin
densities through the shared interaction channels. The internal
cross-spin Hartree term has its own factorization of rank $n_H$, with
factor sizes $K_\uparrow\times n_H$ and $K_\downarrow\times n_H$.
These describe the interaction between the two spin-pair spaces within
the block, separately from the channels carrying fields outside it.
We contract the factors directly, avoiding a dense
$K_\uparrow\times K_\downarrow$ matrix.

Exchange is inexpensive when both the interaction and the state are
compressible. A small MPO rank by itself does not ensure a low-rank
internal pair interaction: even one exponential can give a full-rank
$\boldsymbol\Gamma^\sigma$ for
generic orbitals. Small state rank, on the other hand, immediately
makes the pair space small. The large UHF chains require at most
five active modes per spin, or 15 pairs, all of which we retain.

The density-space compression introduced in
Sec.~\ref{sec:three-compressions} acts directly on this pair space.
As the orbitals change, so do the Fock fields. The retained pair space
must describe these changes as well as the current density. We therefore
target the density together with its operator responses, analogous to
targeting several states in DMRG.

The targets are symmetric $m_\sigma\times m_\sigma$ matrices
$\mathbf A_\mu$: for example, a density matrix in the retained orbital
basis, its change under an orbital rotation, or the field induced by
that change.
We encode each as a length-$K_\sigma$ vector with entries
$y_{\mu,pq}=\sqrt{2-\delta_{pq}}(A_\mu)_{pq}$ for $p\le q$.
The $\sqrt{2}$ on off-diagonal entries makes the vector inner product
equal to the matrix Frobenius inner product, preserving the norm under
orbital rotations. With $\mu=1,\ldots,n_t$ labeling the chosen targets
and nonnegative weights $\omega_\mu$ setting their normalization and relative
importance, form
\begin{equation}
 \begin{aligned}
 \mathbf D_{\rm pair}&=\sum_\mu \omega_\mu\vec y_\mu\vec y_\mu^{\,T}
      =\mathbf Q\mathbf Q^T,\\
 \mathbf Q&=[\sqrt{\omega_1}\vec y_1,\sqrt{\omega_2}\vec y_2,\ldots].
 \end{aligned}
 \label{eq:methods-response-gram}
\end{equation}
Here $\mathbf Q$ is $K_\sigma\times n_t$ and $\mathbf D_{\rm pair}$
is $K_\sigma\times K_\sigma$. The leading eigenvectors of
$\mathbf D_{\rm pair}$ identify a smaller pair basis. When there
are few target vectors, they can be obtained from the
$n_t\times n_t$ Gram matrix $\mathbf Q^T\mathbf Q$.
The positive semidefinite matrix $\mathbf D_{\rm pair}$ plays the role
of a targeting density matrix, but it is neither the physical
two-particle density matrix nor a projector. If the pair coordinates are
divided into two groups, the off-diagonal block does not determine the
weight within either group, so an off-diagonal SVD alone is insufficient.
In our long-chain calculations the pair spaces are already small enough
to retain in full.

For fixed physical block size and one-body bandwidth, the half-sweep
time has the schematic form
\begin{equation}
 T_{1/2}\sim \NH f(\chi,m_\uparrow,m_\downarrow,K_\uparrow,K_\downarrow).
 \label{eq:cost}
\end{equation}
Keeping the channel, state, and pair ranks small makes each sweep inexpensive.
Obtaining a solution efficiently also requires an effective local
update, so that the number of sweeps remains modest.

\section{One HF step at each position}
\label{sec:methods-optimizer}

In DMRG, the center problem is small, but its energy is that of the
full system. The left and right block bases restrict the states
available to the center. With these bases fixed, optimizing the center
is a variational minimization of the full-system energy
\cite{Schollwock2011}.

The same principle applies to HF-DMRG. The center orbitals and stored
block transformations specify a full Slater determinant, with the
completed occupied modes held fixed during the local solve. When we include
their energies and fields and retain all orbital pairs, as in our
calculations, the center functional is exactly the HF energy of this
determinant for the chosen MPO Hamiltonian. Each such energy is an upper
bound to the full-basis RHF or UHF minimum of that Hamiltonian. The block
bases restrict the allowed variations, but the incoming determinant lies
in this space, so any update that lowers the center functional also lowers
the energy of the whole chain before truncation.

We assemble the projection of Eq.~\eqref{eq:fock} directly in the
orthonormal center space of Eq.~\eqref{eq:methods-center-space}.
The matrix combines the projected one-body operator with Hartree fields
from both spins and same-spin exchange, evaluated from the completed
orbitals in the environments and the active center density.
No global Fock matrix is formed. The center holds the
total number of electrons of each spin minus those in the completed
occupied modes of the two environments, just as a DMRG center has a fixed
particle-number sector. Diagonalizing its Fock matrix and filling this
number of lowest levels gives an Aufbau proposal.

We make one HF update per position, allowing the orbitals and their
self-consistent fields to converge together over successive sweeps.
This keeps the local optimization inexpensive and was sufficient
for the hydrogen chains studied here, which converged in only a few
full sweeps. Additional local updates might reduce the number of sweeps
at a greater cost per position; we have not optimized that balance.
More difficult systems may benefit from additional local iterations or
other convergence strategies.

The small HF problem is nonlinear: its Fock matrix depends on the
occupied state. A single Fock diagonalization solves the problem in the current field.
Since the proposed orbitals change that field, we check their full HF
energy before accepting the update. Finite state truncation is part of this
comparison: both the incoming state and the candidate are factored
with the same outgoing selection rule, including the completed-block
contributions. This measures the gain from optimization separately
from the change caused by truncation.

If the full step raises the energy, we try successively shorter
rotations toward the proposed occupied space. To construct these
rotations, we first align the old and new occupied orbitals. An
orthogonal rotation among occupied orbitals changes their gauge but
leaves their projector unchanged, and hence the determinant up to an
overall sign.

For one spin, let $\mathbf C_0$ and $\mathbf C_1$ contain the incoming
and proposed center orbitals. Each has $\dim\mathcal H_{\sigma,a}$ rows
and $n$ orthonormal columns, where $n$ is the number of occupied center
orbitals. We take the SVD of their $n\times n$ overlap matrix,
\begin{equation}
 \mathbf C_0^T\mathbf C_1
   =\mathbf O_0\boldsymbol\Sigma\mathbf O_1^T,
 \qquad \Sigma_{pp}=\cos\theta_p.
 \label{eq:methods-orbital-alignment}
\end{equation}
The diagonal matrix $\boldsymbol\Sigma$ contains the singular values.
The $n\times n$ orthogonal matrices $\mathbf O_0$ and $\mathbf O_1$
align the two occupied bases: the columns of $\mathbf C_0\mathbf O_0$
and $\mathbf C_1\mathbf O_1$ are paired principal vectors
$\vec c_p^{\,0}$ and $\vec c_p^{\,1}$, with overlap $\cos\theta_p$.
For nonorthogonal principal pairs this fixes the alignment independently
of orbital signs and ordering. Exactly orthogonal pairs admit more
than one shortest path. For the chosen alignment the path is
\begin{equation}
 \vec c_p(t)=\vec c_p^{\,0}\cos(t\theta_p)
                 +\vec v_p\sin(t\theta_p),\qquad0\le t\le1,
 \label{eq:methods-geodesic}
\end{equation}
where $p$ labels an occupied orbital and
$\vec v_p=(\vec c_p^{\,1}-\vec c_p^{\,0}\cos\theta_p)/\sin\theta_p$
is the normalized direction orthogonal to the incoming occupied
space. A pair with $\theta_p=0$ is left unchanged. These coefficient
vectors have length $\dim\mathcal H_{\sigma,a}$, the size of the current
center in Eq.~\eqref{eq:methods-center-space}; the spin label is
suppressed in Eq.~\eqref{eq:methods-geodesic}.
This geodesic keeps the orbitals orthonormal and the density matrix
idempotent \cite{EdelmanAriasSmith1998}. UHF uses the same $t$ for both spins, choosing it from
their combined energy. We accept the largest tested step that does
not raise the energy beyond the floating-point allowance; if none
passes, we leave the occupied state unchanged before advancing.

Advancing the sweep also compresses the orbital space according to
the discarded-weight criterion. As in two-site DMRG, this truncation
can slightly raise the energy, even when the orbital optimization
lowers it. We measure the truncation noise by sweeping
without orbital optimization and include it in the convergence checks below.

\begin{table}[tbp]
\caption{\label{tab:h100-ranks}Maximum active-state dimensions $m$ and
pair dimensions $K=m(m+1)/2$ during H$_{100}$ nonlinear calculations
from dimer (RHF) and atomic N\'eel (UHF) starts, with state cutoff
$10^{-12}$. UHF dimensions refer to each spin.}
\begin{ruledtabular}
\begin{tabular}{lccc}
Method & $R$ (bohr) & Maximum $m$ & Maximum $K$ \\
UHF & 3.6 & 5 & 15 \\
RHF & 3.6 & 9 & 45 \\
UHF & 1.25 & 12 & 78
\end{tabular}
\end{ruledtabular}
\end{table}

\section{Accuracy and retained dimensions}
\label{sec:accuracy}

Untruncated RHF and UHF sweeps on H$_2$, H$_{10}$, and H$_{20}$ agree
with dense HF calculations using the same finite Hamiltonian.
We also compare the block and dense formulas on identical determinants,
testing the arithmetic separately from nonlinear convergence. In independent-spin UHF tests,
energy and spin-Fock Frobenius errors are below $1.2\times10^{-13}$ Ha
and $3\times10^{-14}$ Ha, respectively.

For H$_{1000}$, the RHF dimer and UHF atomic N\'eel starts both converge
within five full sweeps. At most nine active modes are needed for RHF
and five per spin for UHF. State compression alone thus reduces the
1000-electron exchange problem to small pair spaces.

\begin{figure}[tb]
\centering
\includegraphics[width=\columnwidth]{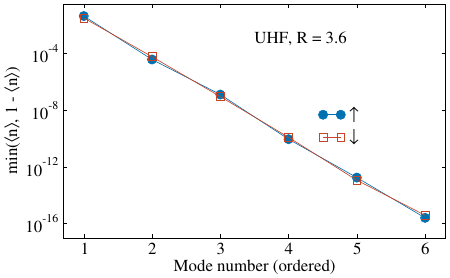}
\caption{\label{fig:restriction}At a central cut of UHF H$_{100}$ ($R=3.6$ bohr), the restricted
occupations $\langle n\rangle$ rapidly approach zero or one,
leaving only a few active modes per spin. The six largest distances
from zero or one are shown in descending order.}
\end{figure}

The smaller UHF rank reflects the physics of the stretched chain.
For H$_{1000}$ its staggered moment, defined as the average alternating
spin-up-minus-spin-down population per atomic interval, is about 0.85.
Its energy is about 55 mHa per atom below RHF, and allowing the spins to
localize on alternate atoms also makes the state more compressible.
We therefore use UHF for the large-chain calculations. At the smaller
spacing $R=1.25$ bohr, the orbitals are more delocalized and more modes
must be retained. Figure~\ref{fig:restriction} shows how rapidly the UHF
occupations approach zero or one in the stretched chain.
Table~\ref{tab:h100-ranks} gives the dimensions for H$_{100}$ at both
spacings. At $R=3.6$ bohr these maxima already match those of
H$_{1000}$, reflecting the local structure at a cut rather than the
total chain length.

\begin{table*}[t]
\caption{\label{tab:stateaccuracy}State-compression accuracy for RHF
H$_{1000}$, with $n_{\rm occ}=500$ doubly occupied spatial orbitals.
Optimization starts from the dimer product state: the reference at
$10^{-12}$ converges in nine half-sweeps; the looser rows show results
after 20. Compression-only rows apply one full sweep to the reference,
without energy minimization. The cutoff bounds the combined discarded
weight at each cut; the measured column gives its maximum in the last
full sweep. Here $m$ is the maximum number of active orbitals in a
final block, and $K=m(m+1)/2$. All energies use the same compressed interaction.
Relative to the reference, $\delta e$ is the energy difference per atom and
$\delta P=\|\mathbf D-\mathbf D_{\rm ref}\|_F/\sqrt{n_{\rm occ}}$
compares the one-particle density matrices for one spin.}
\begin{ruledtabular}
\begin{tabular}{clcccc}
\shortstack{Allowed\\discarded weight} & Calculation & \shortstack{Discarded weight\\(last full sweep)} & Maximum $m/K$ & $\delta e$ (Ha/atom) & $\delta P$ \\
$10^{-6}$ & Optimization & $5.00\times10^{-9}$ & $4/10$ & $1.41\times10^{-6}$ & $1.88\times10^{-3}$ \\
 & Compression only & $7.82\times10^{-7}$ & $4/10$ & $4.01\times10^{-6}$ & $1.61\times10^{-3}$ \\[2pt]
$10^{-8}$ & Optimization & $1.20\times10^{-9}$ & $6/21$ & $6.02\times10^{-9}$ & $1.01\times10^{-4}$ \\
 & Compression only & $6.05\times10^{-9}$ & $6/21$ & $1.42\times10^{-8}$ & $8.40\times10^{-5}$ \\[2pt]
$10^{-10}$ & Optimization & $6.24\times10^{-11}$ & $7/28$ & $2.44\times10^{-10}$ & $2.09\times10^{-5}$ \\
$10^{-12}$ & Optimization & $9.71\times10^{-13}$ & $9/45$ & $0$ & $0$
\end{tabular}
\end{ruledtabular}
\end{table*}

\subsection{State truncation and convergence}

State truncation, incomplete orbital optimization, and interaction
compression introduce different errors. As in DMRG, a sweep can have
nearly stopped lowering the energy while further accuracy is limited
by the retained spaces. Small changes in these spaces can also produce
visible orbital changes with little effect on the energy. We therefore
judge convergence relative to the truncation noise, rather than requiring
identical occupied spaces on successive sweeps. These checks use energy
and local orbital changes, with additional sweeps in both directions,
rather than an independently evaluated full-system orbital residual.
Appendix~\ref{app:convergence}
gives the numerical settings for the long-chain calculations.

Table~\ref{tab:stateaccuracy} illustrates the accuracy gained by retaining
more modes in RHF H$_{1000}$. Six active modes give an energy within
about $10^{-8}$ Ha per atom of the tighter reference, even though
there are 500 occupied spatial orbitals. The compression-only rows
show what happens when that accurate reference is swept at a looser
cutoff without further energy minimization. Compression alone produces
errors comparable to those left after optimization at the same cutoff.

The discarded weight measures the
\emph{new} loss at a cut, not the error already present in the state.
Once compressed, a state can pass through the same restricted spaces
with little further loss. A small discarded weight late in the
calculation can therefore coexist with an appreciable energy error.
Orbital optimization reduces this error within the retained spaces.
Further accuracy requires enlarging those spaces: rebuilding the block
bases of the loosest state at the reference cutoff reduced its energy
error to $9.6\times10^{-12}$ Ha per atom within one full optimization sweep.

\begin{table}[t]
\caption{\label{tab:interactionaccuracy}Effect of interaction compression.
The target error $\epsilon_V$ applies to individual interaction matrix
elements $V_{ij}$ and is measured in Ha. Energy differences below are
per atom: $\delta e_V$ is the compressed-minus-uncompressed
energy of one unchanged UHF H$_{34\,000}$ determinant.
$\delta e_{\rm state}$ compares the uncompressed energies of RHF H$_{1000}$
determinants optimized at each $\epsilon_V$, relative to the
$10^{-8}$ result.}
\begin{ruledtabular}
\begin{tabular}{ccc}
$\epsilon_V$ (Ha) & UHF H$_{34\,000}$ & RHF H$_{1000}$ \\
 & $\delta e_V$ (Ha/atom) & $\delta e_{\rm state}$ (Ha/atom) \\
$10^{-5}$ & $2.45\times10^{-6}$ & $1.24\times10^{-12}$ \\
$10^{-7}$ & $8.62\times10^{-7}$ & $2.16\times10^{-15}$ \\
$10^{-8}$ & $8.70\times10^{-7}$ & $0$
\end{tabular}
\end{ruledtabular}
\end{table}

\begin{figure}[!tb]
\centering
\includegraphics[width=\columnwidth]{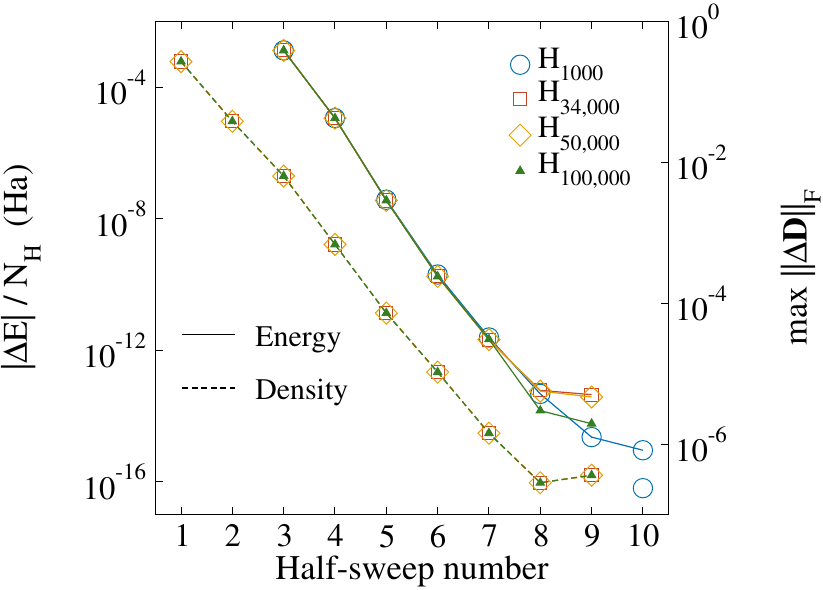}
\caption{\label{fig:convergence}UHF convergence from atomic N\'eel
product states at $R=3.6$ bohr. Convergence changes little over two
orders of magnitude in chain length.
Solid curves (left axis): energy change per atom over one full sweep,
$|\Delta E_\ell|/\NH=|E_\ell-E_{\ell-2}|/\NH$, measured at the end of
half-sweep $\ell$. Dashed curves (right axis): largest Frobenius norm
of the change in a local
spin density matrix from an orbital update, before truncation,
maximized over positions and spins in that half-sweep.
For H$_{1000}$, only the final density-matrix change was recorded,
so that series contains a single point.}
\end{figure}

\begin{figure}[t]
\centering
\includegraphics[width=\columnwidth]{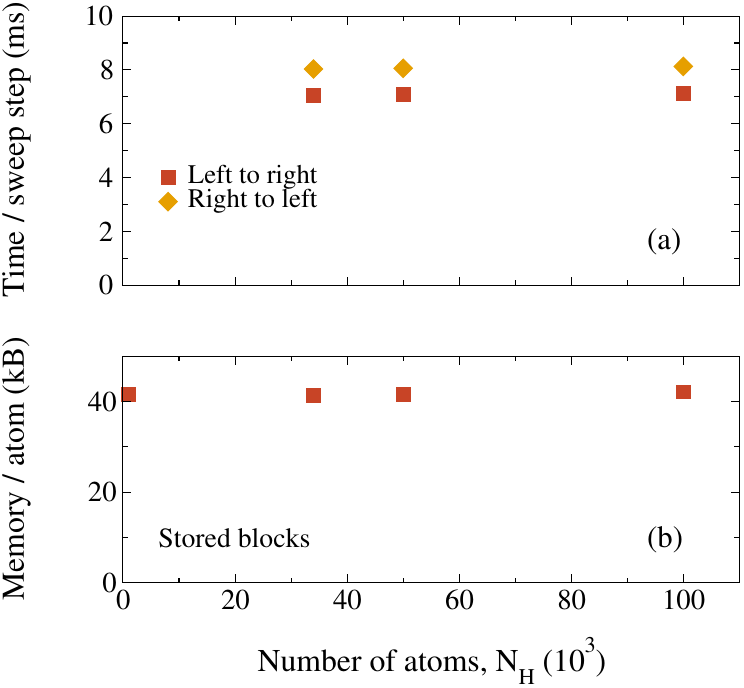}
\caption{\label{fig:scaling}Time and stored memory per atom change little
with chain length.
UHF chains at $R=3.6$ bohr retain at most five active modes per spin.
(a) Half-sweep time divided by $\NH$, taking the median over late
optimization half-sweeps in each direction. Each bulk block
contains 18 spatial basis functions spanning one atomic interval. Each step
advances the two-block center by one block, with $\NH$ steps per half-sweep
in these chains. (b) Stored-block memory per atom in the converged
states. Peak process memory is listed in Table~\ref{tab:benchmarks}.}
\end{figure}

\begin{table*}[t]
\caption{\label{tab:benchmarks}UHF benchmarks on one Apple M4 Pro Mac mini
with 64~GB memory, one Julia thread, and four OpenBLAS threads. Times are in
seconds and memory in decimal GB. Each solution took nine half-sweeps.
Further sweeps check the solution with the direction reversed, the
spins interchanged, and orbital optimization disabled.
The solution subtotal sums the optimization sweeps, initialization, and
interaction construction; checks and other elapsed time are listed separately.
The two largest runs omit the uncompressed-interaction evaluation.
The MPO dimension at a site cut includes channels within an 18-function
block, in addition to those connecting whole blocks.}
\begin{ruledtabular}
\begin{tabular}{lrrr}
Quantity & H$_{34\,000}$ & H$_{50\,000}$ & H$_{100\,000}$ \\
Spatial basis functions & 612,003 & 900,003 & 1,800,003 \\
Maximum MPO dimension (site cut) & 132 & 133 & 136 \\
Optimizing sweeps (s) & 2210.4 & 3252.8 & 6536.0 \\
Initialization (s) & 36.9 & 52.3 & 100.8 \\
Interaction construction (s) & 133.6 & 187.5 & 378.1 \\
Solution subtotal (s) & 2380.9 & 3492.7 & 7014.8 \\
Additional sweeps checking solution (s) & 2279.1 & 3366.8 & 6771.0 \\
Uncompressed energy evaluation (s) & 13430.1 & --- & --- \\
Other elapsed time (s) & 160.7 & 267.7 & 646.6 \\
Total elapsed time (s) & 18250.8 & 7127.1 & 14432.4 \\
Stored blocks (GB) & 1.402 & 2.073 & 4.208 \\
Peak process memory (GB) & 8.17 & 13.05 & 20.83
\end{tabular}
\end{ruledtabular}
\end{table*}

\subsection{Interaction compression and energy evaluation}

An interaction accurate enough to optimize the orbitals can still
give an appreciable offset in their energy. Table~\ref{tab:interactionaccuracy}
separates the two effects: evaluating one determinant with different
interaction approximations, and reoptimizing the determinant before
evaluating its uncompressed energy. The large chains use
$\epsilon_V=10^{-5}$ Ha, which shifts the H$_{34\,000}$ energy by
$2.45\times10^{-6}$ Ha per atom. However, the H$_{1000}$ comparison
shows that tightening the interaction changes the optimized orbitals
so little that their uncompressed energies almost coincide.

A component comparison at H$_{1000}$ shows that the offset is mainly
electrostatic. Small systematic errors in the
interaction tail accumulate in the long-range Hartree sum, whereas
exchange also contains the rapidly decaying density matrix.
At the tightest H$_{1000}$ setting, the dominant contribution comes
from the fitted correction at separations beyond its fitting range.
Reducing $\epsilon_V$ while holding that range fixed therefore does not
necessarily reduce the energy offset, even when individual interaction
entries meet the requested accuracy.

We can remove this offset by evaluating the uncompressed interaction
on the final stored determinant, much as one evaluates a more accurate
Hamiltonian on a converged DMRG state. This streams through the blocks
with modest memory, but remains quadratic in work. We perform this
evaluation through H$_{34\,000}$, obtaining $-16347.012280$ Ha, or
$-0.480794479$ Ha per atom. The larger timing runs report energies
with the compressed interaction. The more accurate expectation value
does not by itself make the orbitals stationary for the uncompressed
Hamiltonian.

\section{Convergence and scaling of the long chains}
\label{sec:convergence}

A linear-cost sweep gives an efficient calculation only if the number
of sweeps also stays small. Starting from atomic N\'eel products, the
chains from H$_{1000}$ to H$_{100\,000}$ converge within five full sweeps,
with nearly the same reduction in energy per atom on each sweep and
similar local density-matrix changes (Fig.~\ref{fig:convergence}).
The local relaxation changes little over two orders of magnitude in length.

The simplicity of this convergence has a physical origin. The start
already has local charge neutrality and the correct short-range spin
pattern. The main task is to adjust nearby orbitals, not to transport
charge across the chain. This is a favorable test of local relaxation;
long-wavelength charge redistribution could require many more sweeps.

For these chains, a single HF iteration at each sweep position---building
and diagonalizing the local Fock matrices---is enough: every full orbital
update is accepted in the large calculations. Less favorable starts in
the small-chain tests sometimes require the partial orbital rotations
described in Sec.~\ref{sec:methods-optimizer}.

The retained dimensions are equally stable with length. Four or five
active modes per spin suffice, leaving only 10--15 orbital pairs for
exchange. The interaction channel count grows only slightly as
additional exponentials describe the longer Coulomb tail. Thus
increasing the chain length mainly adds more steps, not larger local
problems.

Table~\ref{tab:benchmarks} gives timings on a 64-GB Apple M4 Pro
Mac mini, using one Julia thread and four OpenBLAS threads for local
linear algebra. Sweep positions are processed sequentially.
Obtaining the H$_{100\,000}$ solution takes about two hours,
including initialization and interaction construction. Further sweeps
checking the solution roughly double that time. The separate
uncompressed energy evaluation dominates the H$_{34\,000}$ run.

Doubling the length from H$_{50\,000}$ to H$_{100\,000}$ doubles
the solution time to within one percent. Figure~\ref{fig:scaling} shows why:
optimization sweeps cost about 7--8 ms per sweep step, almost independent
of length.
Together with the unchanged number of sweeps, this gives nearly
proportional time to solution for this family, including both Hartree
and exchange.

Storage behaves similarly: the memory per atom in the stored blocks
is nearly constant, while the small moving center state does not
grow with length. For H$_{100\,000}$, global occupied-orbital coefficients
would require about 1.44 TB for the two spins in double precision;
the stored blocks use 4.2 GB. Neither initialization nor later sweeps
construct those global coefficients. With work arrays and all other
process memory included, the calculation uses about 21 GB, less than
one third of the Mac mini's memory.

\begin{figure}[t]
\centering
\includegraphics[width=\columnwidth]{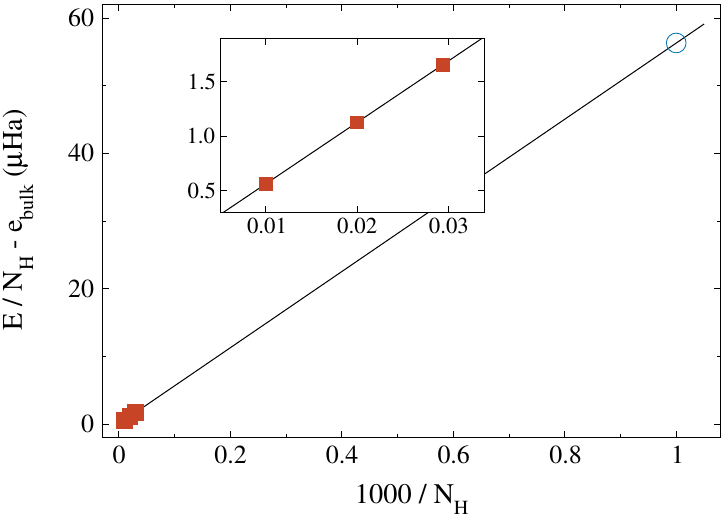}
\caption{\label{fig:energyfit}For UHF chains at $R=3.6$ bohr, the nearly constant energy of the two
ends, $e_{\rm end}$, gives the per-atom correction
$E/\NH-e_{\rm bulk}\simeq e_{\rm end}/\NH$ fitted by the line.
Squares show H$_{34\,000}$, H$_{50\,000}$,
and H$_{100\,000}$, enlarged in the inset. The circle is H$_{1000}$,
excluded from the fit because its interaction approximation uses fewer
exponentials for the Coulomb tail. All energies use the compressed
interaction; fit residuals are of order $10^{-10}$ Ha per atom.}
\end{figure}

The long chains also provide a check on the size dependence of the energy.
For a uniform chain it has an extensive bulk term
and a nearly constant contribution from the two ends,
$E(\NH)\simeq\NH e_{\rm bulk}+e_{\rm end}$.
Figure~\ref{fig:energyfit} plots the end contribution per atom,
$e_{\rm end}/\NH$, using energies for the compressed interaction.
The three largest chains follow this form to about $10^{-10}$ Ha per
atom, much smaller than the interaction-compression offset.
Appendix~\ref{app:size-consistency} gives the fitted values.

\section{Dimensionality and extensions}
\label{sec:limits}

The hydrogen chain is an almost ideal case for HF-DMRG, for much the
same reason that a gapped chain is ideal for ordinary DMRG. The
important quantity is how much state information crosses a cut,
not how many electrons lie on either side. In a long, narrow system
only a few occupied modes may be shared across each cut. Increasing
the length then adds more blocks without enlarging the local problem.
For a strip of fixed width the same reasoning can apply, at a larger
cost set by the width. In two or three dimensions at fixed aspect
ratio, the cut crosses a growing boundary and usually requires more
modes. Gapless states or a poor ordering can also increase the ranks.

Although the one-particle formulation avoids the exponential number
of many-body occupation patterns, its pair dimension grows as $m^2$
and subsequent matrix operations can be more costly still. A growing
boundary can therefore make HF-DMRG expensive well before the number
of electrons becomes a limitation. The restriction spectrum remains
the natural guide to feasibility.

Other systems will favor a different balance of the three
compressions. For these chains, state compression leaves only 10--15
pairs per spin, making further pair-response compression unnecessary.
In a very large basis with few electrons, the entire occupied space
may already be modest while the interaction and response spaces
dominate. A compact interaction can also accommodate disorder:
exponential channels propagate with bond-dependent factors rather than
a common unit-cell transfer. More general interactions can be carried
by factors constructed separately at each cut.

Inexpensive steps do not guarantee rapid convergence. Long-range
charge transfer and competing spin patterns may require many more
sweeps than the local relaxation studied here. As in conventional
HF, locally lowering the energy does not guarantee the global minimum.
Large tight-binding and moir\'e systems are natural applications
where an HF calculation can be useful even when correlated DMRG is
out of reach. Local and semilocal Kohn--Sham theory has the same
basic need to update a state-dependent field \cite{KohnSham1965};
hybrid functionals also contain the exchange problem treated here.
We will describe general dense and sliced interaction representations,
gausslet chemistry applications, and history-based acceleration for
their more demanding convergence problems in a subsequent paper.

\section{Conclusions}
\label{sec:conclusions}

HF-DMRG applies the familiar finite-system DMRG sweep to a single
Slater determinant. We remove filled and empty block modes from
the repeated local work, incorporate their fields into the environments,
and save their transformations for the return sweep. Compressing the
density--density interaction gives a compact
description of long-range fields. Separate spin spaces keep exchange
inexpensive.

For uniform antiferromagnetic hydrogen chains, this combination gives
nearly linear solution time through 100,000 electrons in 1.8 million
spatial basis functions. Both the local dimensions and the number of
sweeps stay small. The largest chain is initialized and converged in
about two hours, using about 21 GB on a 64-GB Mac mini. A more
accurate final energy can be evaluated from the same block state
with the uncompressed interaction; this separate quadratic calculation
is performed through 34,000 atoms. The result illustrates how DMRG's
way of organizing states, operators, and local optimization can be
valuable well beyond correlated many-body wavefunctions.

An open-source implementation of these algorithms is available in the
\texttt{HFDMRG.jl} Julia package \cite{HFDMRGSoftware}.

\begin{acknowledgments}
This work was supported by the U.S. National Science Foundation under
Grant DMR-2412638. The author thanks OpenAI for providing research access
to ChatGPT and Codex, and Anthropic for discounted access to Claude.
A public implementation is available at
\url{https://github.com/srwhite59/HFDMRG.jl}.
The author has no conflicts to disclose.

An earlier HF-DMRG implementation, without MPO compression, had been
used by the author for about five years. Development of the present
method used OpenAI's ChatGPT and Codex (Sol and Astra) extensively for
algorithmic discussions, programming, testing, analysis, and manuscript
and figure preparation. Anthropic's Claude (Fable) assisted with critical
review of the algorithms, software, and manuscript, and with diagram
development. The author directed the work and is responsible for its
scientific content. Numerical checks are described in
Sec.~\ref{sec:accuracy}.
\end{acknowledgments}

\appendix
\raggedbottom
\section{Sequential cross-cut factorization}
\label{app:sequential-svd}

Here we construct the source, transfer, and sink matrices of the
interaction MPO by successive SVDs. The purpose is to obtain compatible
channel bases at neighboring cuts without storing the long singular-vector
matrices of every full cross-cut block. We also show why each smaller
SVD is equivalent to a cross-cut SVD before truncation.

Divide the sites into ordered blocks $B_1,\ldots,B_{n_b}$, with $d_a$
sites in $B_a$. In subscripts on $\mathbf V$, $<a$ denotes all sites in blocks
before $B_a$, $>a$ all sites after it, and $\ge a$ includes $B_a$
and all later blocks. Suppose the previous step has given
\begin{equation}
 \mathbf V_{<a,\ge a}\simeq\mathbf L_{a-1}\mathbf X_{a-1},
 \qquad\mathbf L_{a-1}^T\mathbf L_{a-1}=\mathbf I_{\chi_{a-1}}.
 \label{eq:app-incoming-cut}
\end{equation}
The $\chi_{a-1}$ columns of $\mathbf L_{a-1}$ are an orthonormal
basis for the retained left interaction patterns. The columns of
$\mathbf X_{a-1}$ still correspond to individual sites in $B_a$ and all later
blocks. With $N_L=\sum_{c\le a}d_c$ and $N_R=\sum_{c>a}d_c$ at the
current cut, these two factors have sizes
$(N_L-d_a)\times\chi_{a-1}$ and $\chi_{a-1}\times(d_a+N_R)$.
First separate the $d_a$ columns belonging to $B_a$:
\begin{equation}
 \mathbf X_{a-1}=[\,\bar{\mathbf S}_a\mid\mathbf X_{a-1}^{\rm tail}\,].
 \label{eq:app-sink-definition}
\end{equation}
This defines the $\chi_{a-1}\times d_a$ sink $\bar{\mathbf S}_a$.
Interactions arriving from earlier blocks end on $B_a$ through this
matrix. The remaining $\chi_{a-1}\times N_R$ matrix
$\mathbf X_{a-1}^{\rm tail}$ describes interactions continuing to later blocks.

Stack the new physical interaction rows above that remaining factor:
\begin{equation}
 \mathbf A_a=\begin{pmatrix}
 \mathbf V_{a,>a}\\\mathbf X_{a-1}^{\rm tail}
 \end{pmatrix}.
 \label{eq:app-svd-input}
\end{equation}
Here $\mathbf V_{a,>a}$ has rows in $B_a$ and columns in all later
blocks, so its size is $d_a\times N_R$. The stacked matrix
$\mathbf A_a$ has size $(d_a+\chi_{a-1})\times N_R$.
Perform its thin SVD and retain $\chi_a$ singular directions:
\begin{equation}
 \mathbf A_a\simeq\mathbf U_a\boldsymbol\Sigma_a\mathbf W_a^T
       \equiv\mathbf U_a\mathbf X_a,
 \label{eq:methods-sequential-svd}
\end{equation}
where $\boldsymbol\Sigma_a$ is the $\chi_a\times\chi_a$ diagonal matrix
of retained singular values, and $\mathbf U_a$ and $\mathbf W_a$ contain
the corresponding left and right singular vectors. Their sizes are
$(d_a+\chi_{a-1})\times\chi_a$ and $N_R\times\chi_a$.
Thus $\mathbf X_a=\boldsymbol\Sigma_a\mathbf W_a^T$, of size
$\chi_a\times N_R$, is the right factor passed to the next step.
Splitting $\mathbf U_a$ into the same two row groups as $\mathbf A_a$ defines
the source and transfer:
\begin{equation}
 \mathbf U_a=\begin{pmatrix}\mathbf S_a^T\\\mathbf T_a\end{pmatrix},
 \qquad\mathbf S_a\mathbf S_a^T+\mathbf T_a^T\mathbf T_a=\mathbf I_{\chi_a}.
 \label{eq:app-source-transfer-definition}
\end{equation}
The source $\mathbf S_a^T$ has size $d_a\times\chi_a$ and starts interactions
on $B_a$; the transfer $\mathbf T_a$, of size $\chi_{a-1}\times\chi_a$,
continues incoming channels through the block. In particular,
\begin{equation}
 \mathbf V_{a,>a}\simeq\mathbf S_a^T\mathbf X_a,
 \qquad\mathbf X_{a-1}^{\rm tail}\simeq\mathbf T_a\mathbf X_a.
 \label{eq:app-channel-recurrence}
\end{equation}
Applying the second relation through intervening blocks, then taking
the sink columns at the final block, gives the source--transfer--sink
product in Eq.~\eqref{eq:methods-source-transfer-sink}.

At $B_1$ there are no incoming channels: $\chi_0=0$ and
$\mathbf A_1=\mathbf V_{1,>1}$. Its retained left singular vectors give
$\mathbf S_1^T$.
At the last block, all columns of $\mathbf X_{n_b-1}$ belong to $B_{n_b}$,
so they give $\bar{\mathbf S}_{n_b}$ and no further SVD is needed. There are
no outgoing channels, $\chi_{n_b}=0$. The diagonal blocks $\mathbf V_{aa}$
are kept separately throughout.

To relate this construction to the full cross-cut SVD, restore the
physical rows of the preceding blocks in Eq.~\eqref{eq:app-svd-input}:
\begin{equation}
 \mathbf V_{\le a,>a}\simeq
 \begin{pmatrix}\mathbf 0&\mathbf L_{a-1}\\\mathbf I_{d_a}&\mathbf 0\end{pmatrix}
 \mathbf A_a.
 \label{eq:app-cross-cut-isometry}
\end{equation}
Here $\le a$ includes all blocks through $B_a$. The matrix multiplying
$\mathbf A_a$ has size $N_L\times(d_a+\chi_{a-1})$ and orthonormal
columns, so it preserves its nonzero singular
values. Without previous truncations, these are exactly the singular
values of the original cross-cut block. With previous truncations,
they are those of the interaction carried forward from the earlier
steps. The updated left basis is
\begin{equation}
 \mathbf L_a=
 \begin{pmatrix}\mathbf L_{a-1}\mathbf T_a\\\mathbf S_a^T\end{pmatrix},
 \qquad\mathbf L_a^T\mathbf L_a=\mathbf I_{\chi_a}.
 \label{eq:app-updated-left-basis}
\end{equation}
The updated $\mathbf L_a$ has size $N_L\times\chi_a$. This also identifies
$\mathbf T_a$ as the change of channel basis on the previous blocks.
The long basis $\mathbf L_a$ gives the physical-site interpretation of
the retained channels. During construction, the temporary factor
$\mathbf X_a$ has $\chi_a$ rows and $N_R=\sum_{c>a}d_c$ columns.
The finished representation contains
only $\mathbf V_{aa}$, $\mathbf S_a^T$, $\mathbf T_a$, and
$\bar{\mathbf S}_a$ at each block. Their
sizes depend on the block sizes and channel ranks, not on the number
of sites on either side of the cut.

\section{Bulk and end contributions to the energy}
\label{app:size-consistency}
The fit in Fig.~\ref{fig:energyfit} gives
$e_{\rm bulk}=-0.480793685$ Ha per atom and $e_{\rm end}=0.0562864$ Ha
for both ends together, with residuals of order $10^{-10}$ Ha per atom.
It uses the compressed-interaction energies of the three largest chains.
H$_{1000}$ is shown for comparison but excluded from this fit because
its interaction approximation uses fewer exponentials to represent the
Coulomb tail over the smaller range of separations.

\section{Convergence checks with state truncation}
\label{app:convergence}

The long-chain calculations use state cutoff $10^{-12}$, a limit of 16
active modes per spin, and target interaction-matrix error
$\epsilon_V=10^{-5}$ Ha. The active-mode limit never binds; the largest
spaces contain five modes per spin. Convergence requires small energy changes over
a full sweep in both directions. A floating-point allowance for the
extensive energy supplements the nominal $10^{-11}$ Ha tolerance,
giving thresholds of order $10^{-8}$ Ha for the largest chains.

Further sweeps reverse the direction and interchange the spins.
In the final checking half-sweeps, the largest local spin-density change
from an orbital update, measured in the Frobenius norm as in
Fig.~\ref{fig:convergence}, is compared with $3\times10^{-6}$ plus the
larger of the two spin allowances $\sum_a\sqrt{2w_{\sigma,a}}$.
Here $w_{\sigma,a}$ is the discarded weight of a stored link.
Sweeps without orbital optimization separately measure the energy
change from compression alone.

\bibliographystyle{apsrev4-2}
\bibliography{ref}

\begin{thebibliography}{31}%
\makeatletter
\providecommand \@ifxundefined [1]{%
 \@ifx{#1\undefined}
}%
\providecommand \@ifnum [1]{%
 \ifnum #1\expandafter \@firstoftwo
 \else \expandafter \@secondoftwo
 \fi
}%
\providecommand \@ifx [1]{%
 \ifx #1\expandafter \@firstoftwo
 \else \expandafter \@secondoftwo
 \fi
}%
\providecommand \natexlab [1]{#1}%
\providecommand \enquote  [1]{``#1''}%
\providecommand \bibnamefont  [1]{#1}%
\providecommand \bibfnamefont [1]{#1}%
\providecommand \citenamefont [1]{#1}%
\providecommand \href@noop [0]{\@secondoftwo}%
\providecommand \href [0]{\begingroup \@sanitize@url \@href}%
\providecommand \@href[1]{\@@startlink{#1}\@@href}%
\providecommand \@@href[1]{\endgroup#1\@@endlink}%
\providecommand \@sanitize@url [0]{\catcode `\\12\catcode `\$12\catcode
  `\&12\catcode `\#12\catcode `\^12\catcode `\_12\catcode `\%12\relax}%
\providecommand \@@startlink[1]{}%
\providecommand \@@endlink[0]{}%
\providecommand \url  [0]{\begingroup\@sanitize@url \@url }%
\providecommand \@url [1]{\endgroup\@href {#1}{\urlprefix }}%
\providecommand \urlprefix  [0]{URL }%
\providecommand \Eprint [0]{\href }%
\providecommand \doibase [0]{https://doi.org/}%
\providecommand \selectlanguage [0]{\@gobble}%
\providecommand \bibinfo  [0]{\@secondoftwo}%
\providecommand \bibfield  [0]{\@secondoftwo}%
\providecommand \translation [1]{[#1]}%
\providecommand \BibitemOpen [0]{}%
\providecommand \bibitemStop [0]{}%
\providecommand \bibitemNoStop [0]{.\EOS\space}%
\providecommand \EOS [0]{\spacefactor3000\relax}%
\providecommand \BibitemShut  [1]{\csname bibitem#1\endcsname}%
\let\auto@bib@innerbib\@empty
\bibitem [{\citenamefont {Kohn}\ and\ \citenamefont
  {Sham}(1965)}]{KohnSham1965}%
  \BibitemOpen
  \bibfield  {author} {\bibinfo {author} {\bibfnamefont {W.}~\bibnamefont
  {Kohn}}\ and\ \bibinfo {author} {\bibfnamefont {L.~J.}\ \bibnamefont
  {Sham}},\ }\href {https://doi.org/10.1103/PhysRev.140.A1133} {\bibfield
  {journal} {\bibinfo  {journal} {Phys. Rev.}\ }\textbf {\bibinfo {volume}
  {140}},\ \bibinfo {pages} {A1133} (\bibinfo {year} {1965})}\BibitemShut
  {NoStop}%
\bibitem [{\citenamefont {Goedecker}(1999)}]{Goedecker1999}%
  \BibitemOpen
  \bibfield  {author} {\bibinfo {author} {\bibfnamefont {S.}~\bibnamefont
  {Goedecker}},\ }\href {https://doi.org/10.1103/RevModPhys.71.1085} {\bibfield
   {journal} {\bibinfo  {journal} {Rev. Mod. Phys.}\ }\textbf {\bibinfo
  {volume} {71}},\ \bibinfo {pages} {1085} (\bibinfo {year}
  {1999})}\BibitemShut {NoStop}%
\bibitem [{\citenamefont {Prodan}\ and\ \citenamefont
  {Kohn}(2005)}]{ProdanKohn2005}%
  \BibitemOpen
  \bibfield  {author} {\bibinfo {author} {\bibfnamefont {E.}~\bibnamefont
  {Prodan}}\ and\ \bibinfo {author} {\bibfnamefont {W.}~\bibnamefont {Kohn}},\
  }\href {https://doi.org/10.1073/pnas.0505436102} {\bibfield  {journal}
  {\bibinfo  {journal} {Proc. Natl. Acad. Sci. U.S.A.}\ }\textbf {\bibinfo
  {volume} {102}},\ \bibinfo {pages} {11635} (\bibinfo {year}
  {2005})}\BibitemShut {NoStop}%
\bibitem [{\citenamefont {Li}\ \emph {et~al.}(1993)\citenamefont {Li},
  \citenamefont {Nunes},\ and\ \citenamefont
  {Vanderbilt}}]{LiNunesVanderbilt1993}%
  \BibitemOpen
  \bibfield  {author} {\bibinfo {author} {\bibfnamefont {X.-P.}\ \bibnamefont
  {Li}}, \bibinfo {author} {\bibfnamefont {R.~W.}\ \bibnamefont {Nunes}},\ and\
  \bibinfo {author} {\bibfnamefont {D.}~\bibnamefont {Vanderbilt}},\ }\href
  {https://doi.org/10.1103/PhysRevB.47.10891} {\bibfield  {journal} {\bibinfo
  {journal} {Phys. Rev. B}\ }\textbf {\bibinfo {volume} {47}},\ \bibinfo
  {pages} {10891} (\bibinfo {year} {1993})}\BibitemShut {NoStop}%
\bibitem [{\citenamefont {Niklasson}(2002)}]{Niklasson2002}%
  \BibitemOpen
  \bibfield  {author} {\bibinfo {author} {\bibfnamefont {A.~M.~N.}\
  \bibnamefont {Niklasson}},\ }\href
  {https://doi.org/10.1103/PhysRevB.66.155115} {\bibfield  {journal} {\bibinfo
  {journal} {Phys. Rev. B}\ }\textbf {\bibinfo {volume} {66}},\ \bibinfo
  {pages} {155115} (\bibinfo {year} {2002})}\BibitemShut {NoStop}%
\bibitem [{\citenamefont {Mauri}\ \emph {et~al.}(1993)\citenamefont {Mauri},
  \citenamefont {Galli},\ and\ \citenamefont {Car}}]{MauriGalliCar1993}%
  \BibitemOpen
  \bibfield  {author} {\bibinfo {author} {\bibfnamefont {F.}~\bibnamefont
  {Mauri}}, \bibinfo {author} {\bibfnamefont {G.}~\bibnamefont {Galli}},\ and\
  \bibinfo {author} {\bibfnamefont {R.}~\bibnamefont {Car}},\ }\href
  {https://doi.org/10.1103/PhysRevB.47.9973} {\bibfield  {journal} {\bibinfo
  {journal} {Phys. Rev. B}\ }\textbf {\bibinfo {volume} {47}},\ \bibinfo
  {pages} {9973} (\bibinfo {year} {1993})}\BibitemShut {NoStop}%
\bibitem [{\citenamefont {Yang}(1991)}]{Yang1991}%
  \BibitemOpen
  \bibfield  {author} {\bibinfo {author} {\bibfnamefont {W.}~\bibnamefont
  {Yang}},\ }\href {https://doi.org/10.1103/PhysRevLett.66.1438} {\bibfield
  {journal} {\bibinfo  {journal} {Phys. Rev. Lett.}\ }\textbf {\bibinfo
  {volume} {66}},\ \bibinfo {pages} {1438} (\bibinfo {year}
  {1991})}\BibitemShut {NoStop}%
\bibitem [{\citenamefont {White}\ \emph {et~al.}(1994)\citenamefont {White},
  \citenamefont {Johnson}, \citenamefont {Gill},\ and\ \citenamefont
  {Head-Gordon}}]{WhiteCFMM1994}%
  \BibitemOpen
  \bibfield  {author} {\bibinfo {author} {\bibfnamefont {C.~A.}\ \bibnamefont
  {White}}, \bibinfo {author} {\bibfnamefont {B.~G.}\ \bibnamefont {Johnson}},
  \bibinfo {author} {\bibfnamefont {P.~M.~W.}\ \bibnamefont {Gill}},\ and\
  \bibinfo {author} {\bibfnamefont {M.}~\bibnamefont {Head-Gordon}},\ }\href
  {https://doi.org/10.1016/0009-2614(94)01128-1} {\bibfield  {journal}
  {\bibinfo  {journal} {Chem. Phys. Lett.}\ }\textbf {\bibinfo {volume}
  {230}},\ \bibinfo {pages} {8} (\bibinfo {year} {1994})}\BibitemShut {NoStop}%
\bibitem [{\citenamefont {Schwegler}\ \emph {et~al.}(1997)\citenamefont
  {Schwegler}, \citenamefont {Challacombe},\ and\ \citenamefont
  {Head-Gordon}}]{Schwegler1997}%
  \BibitemOpen
  \bibfield  {author} {\bibinfo {author} {\bibfnamefont {E.}~\bibnamefont
  {Schwegler}}, \bibinfo {author} {\bibfnamefont {M.}~\bibnamefont
  {Challacombe}},\ and\ \bibinfo {author} {\bibfnamefont {M.}~\bibnamefont
  {Head-Gordon}},\ }\href {https://doi.org/10.1063/1.473833} {\bibfield
  {journal} {\bibinfo  {journal} {J. Chem. Phys.}\ }\textbf {\bibinfo {volume}
  {106}},\ \bibinfo {pages} {9708} (\bibinfo {year} {1997})}\BibitemShut
  {NoStop}%
\bibitem [{\citenamefont {K{\"o}ppl}\ and\ \citenamefont
  {Werner}(2016)}]{KopplWerner2016}%
  \BibitemOpen
  \bibfield  {author} {\bibinfo {author} {\bibfnamefont {C.}~\bibnamefont
  {K{\"o}ppl}}\ and\ \bibinfo {author} {\bibfnamefont {H.-J.}\ \bibnamefont
  {Werner}},\ }\href {https://doi.org/10.1021/acs.jctc.6b00251} {\bibfield
  {journal} {\bibinfo  {journal} {J. Chem. Theory Comput.}\ }\textbf {\bibinfo
  {volume} {12}},\ \bibinfo {pages} {3122} (\bibinfo {year}
  {2016})}\BibitemShut {NoStop}%
\bibitem [{\citenamefont {Sa{\l}ek}\ \emph {et~al.}(2007)\citenamefont
  {Sa{\l}ek}, \citenamefont {H{\o}st}, \citenamefont {Th{\o}gersen},
  \citenamefont {J{\o}rgensen}, \citenamefont {Manninen}, \citenamefont
  {Olsen}, \citenamefont {Jans{\'i}k}, \citenamefont {Reine}, \citenamefont
  {Paw{\l}owski}, \citenamefont {Tellgren}, \citenamefont {Helgaker},\ and\
  \citenamefont {Coriani}}]{Salek2007}%
  \BibitemOpen
  \bibfield  {author} {\bibinfo {author} {\bibfnamefont {P.}~\bibnamefont
  {Sa{\l}ek}}, \bibinfo {author} {\bibfnamefont {S.}~\bibnamefont {H{\o}st}},
  \bibinfo {author} {\bibfnamefont {L.}~\bibnamefont {Th{\o}gersen}}, \bibinfo
  {author} {\bibfnamefont {P.}~\bibnamefont {J{\o}rgensen}}, \bibinfo {author}
  {\bibfnamefont {P.}~\bibnamefont {Manninen}}, \bibinfo {author}
  {\bibfnamefont {J.}~\bibnamefont {Olsen}}, \bibinfo {author} {\bibfnamefont
  {B.}~\bibnamefont {Jans{\'i}k}}, \bibinfo {author} {\bibfnamefont
  {S.}~\bibnamefont {Reine}}, \bibinfo {author} {\bibfnamefont
  {F.}~\bibnamefont {Paw{\l}owski}}, \bibinfo {author} {\bibfnamefont
  {E.}~\bibnamefont {Tellgren}}, \bibinfo {author} {\bibfnamefont
  {T.}~\bibnamefont {Helgaker}},\ and\ \bibinfo {author} {\bibfnamefont
  {S.}~\bibnamefont {Coriani}},\ }\href {https://doi.org/10.1063/1.2464111}
  {\bibfield  {journal} {\bibinfo  {journal} {J. Chem. Phys.}\ }\textbf
  {\bibinfo {volume} {126}},\ \bibinfo {pages} {114110} (\bibinfo {year}
  {2007})}\BibitemShut {NoStop}%
\bibitem [{\citenamefont {White}(1992)}]{White1992}%
  \BibitemOpen
  \bibfield  {author} {\bibinfo {author} {\bibfnamefont {S.~R.}\ \bibnamefont
  {White}},\ }\href {https://doi.org/10.1103/PhysRevLett.69.2863} {\bibfield
  {journal} {\bibinfo  {journal} {Phys. Rev. Lett.}\ }\textbf {\bibinfo
  {volume} {69}},\ \bibinfo {pages} {2863} (\bibinfo {year}
  {1992})}\BibitemShut {NoStop}%
\bibitem [{\citenamefont {White}(1993)}]{White1993}%
  \BibitemOpen
  \bibfield  {author} {\bibinfo {author} {\bibfnamefont {S.~R.}\ \bibnamefont
  {White}},\ }\href {https://doi.org/10.1103/PhysRevB.48.10345} {\bibfield
  {journal} {\bibinfo  {journal} {Phys. Rev. B}\ }\textbf {\bibinfo {volume}
  {48}},\ \bibinfo {pages} {10345} (\bibinfo {year} {1993})}\BibitemShut
  {NoStop}%
\bibitem [{\citenamefont {Schollw{\"o}ck}(2011)}]{Schollwock2011}%
  \BibitemOpen
  \bibfield  {author} {\bibinfo {author} {\bibfnamefont {U.}~\bibnamefont
  {Schollw{\"o}ck}},\ }\href {https://doi.org/10.1016/j.aop.2010.09.012}
  {\bibfield  {journal} {\bibinfo  {journal} {Ann. Phys.}\ }\textbf {\bibinfo
  {volume} {326}},\ \bibinfo {pages} {96} (\bibinfo {year} {2011})}\BibitemShut
  {NoStop}%
\bibitem [{\citenamefont {{\"O}stlund}\ and\ \citenamefont
  {Rommer}(1995)}]{OstlundRommer1995}%
  \BibitemOpen
  \bibfield  {author} {\bibinfo {author} {\bibfnamefont {S.}~\bibnamefont
  {{\"O}stlund}}\ and\ \bibinfo {author} {\bibfnamefont {S.}~\bibnamefont
  {Rommer}},\ }\href {https://doi.org/10.1103/PhysRevLett.75.3537} {\bibfield
  {journal} {\bibinfo  {journal} {Phys. Rev. Lett.}\ }\textbf {\bibinfo
  {volume} {75}},\ \bibinfo {pages} {3537} (\bibinfo {year}
  {1995})}\BibitemShut {NoStop}%
\bibitem [{\citenamefont {White}\ and\ \citenamefont
  {Noack}(1992)}]{WhiteNoack1992}%
  \BibitemOpen
  \bibfield  {author} {\bibinfo {author} {\bibfnamefont {S.~R.}\ \bibnamefont
  {White}}\ and\ \bibinfo {author} {\bibfnamefont {R.~M.}\ \bibnamefont
  {Noack}},\ }\href {https://doi.org/10.1103/PhysRevLett.68.3487} {\bibfield
  {journal} {\bibinfo  {journal} {Phys. Rev. Lett.}\ }\textbf {\bibinfo
  {volume} {68}},\ \bibinfo {pages} {3487} (\bibinfo {year}
  {1992})}\BibitemShut {NoStop}%
\bibitem [{\citenamefont {Chung}\ and\ \citenamefont
  {Peschel}(2001)}]{ChungPeschel2001}%
  \BibitemOpen
  \bibfield  {author} {\bibinfo {author} {\bibfnamefont {M.-C.}\ \bibnamefont
  {Chung}}\ and\ \bibinfo {author} {\bibfnamefont {I.}~\bibnamefont
  {Peschel}},\ }\href {https://doi.org/10.1103/PhysRevB.64.064412} {\bibfield
  {journal} {\bibinfo  {journal} {Phys. Rev. B}\ }\textbf {\bibinfo {volume}
  {64}},\ \bibinfo {pages} {064412} (\bibinfo {year} {2001})}\BibitemShut
  {NoStop}%
\bibitem [{\citenamefont {Peschel}(2003)}]{Peschel2003}%
  \BibitemOpen
  \bibfield  {author} {\bibinfo {author} {\bibfnamefont {I.}~\bibnamefont
  {Peschel}},\ }\href {https://doi.org/10.1088/0305-4470/36/14/101} {\bibfield
  {journal} {\bibinfo  {journal} {J. Phys. A: Math. Gen.}\ }\textbf {\bibinfo
  {volume} {36}},\ \bibinfo {pages} {L205} (\bibinfo {year}
  {2003})}\BibitemShut {NoStop}%
\bibitem [{\citenamefont {Cheong}\ and\ \citenamefont
  {Henley}(2004)}]{CheongHenley2004}%
  \BibitemOpen
  \bibfield  {author} {\bibinfo {author} {\bibfnamefont {S.-A.}\ \bibnamefont
  {Cheong}}\ and\ \bibinfo {author} {\bibfnamefont {C.~L.}\ \bibnamefont
  {Henley}},\ }\href {https://doi.org/10.1103/PhysRevB.69.075111} {\bibfield
  {journal} {\bibinfo  {journal} {Phys. Rev. B}\ }\textbf {\bibinfo {volume}
  {69}},\ \bibinfo {pages} {075111} (\bibinfo {year} {2004})}\BibitemShut
  {NoStop}%
\bibitem [{\citenamefont {Botero}\ and\ \citenamefont
  {Reznik}(2004)}]{BoteroReznik2004}%
  \BibitemOpen
  \bibfield  {author} {\bibinfo {author} {\bibfnamefont {A.}~\bibnamefont
  {Botero}}\ and\ \bibinfo {author} {\bibfnamefont {B.}~\bibnamefont
  {Reznik}},\ }\href {https://doi.org/10.1016/j.physleta.2004.08.037}
  {\bibfield  {journal} {\bibinfo  {journal} {Phys. Lett. A}\ }\textbf
  {\bibinfo {volume} {331}},\ \bibinfo {pages} {39} (\bibinfo {year}
  {2004})}\BibitemShut {NoStop}%
\bibitem [{\citenamefont {Fishman}\ and\ \citenamefont
  {White}(2015)}]{FishmanWhite2015}%
  \BibitemOpen
  \bibfield  {author} {\bibinfo {author} {\bibfnamefont {M.~T.}\ \bibnamefont
  {Fishman}}\ and\ \bibinfo {author} {\bibfnamefont {S.~R.}\ \bibnamefont
  {White}},\ }\href {https://doi.org/10.1103/PhysRevB.92.075132} {\bibfield
  {journal} {\bibinfo  {journal} {Phys. Rev. B}\ }\textbf {\bibinfo {volume}
  {92}},\ \bibinfo {pages} {075132} (\bibinfo {year} {2015})}\BibitemShut
  {NoStop}%
\bibitem [{\citenamefont {Schuch}\ and\ \citenamefont
  {Bauer}(2019)}]{SchuchBauer2019}%
  \BibitemOpen
  \bibfield  {author} {\bibinfo {author} {\bibfnamefont {N.}~\bibnamefont
  {Schuch}}\ and\ \bibinfo {author} {\bibfnamefont {B.}~\bibnamefont {Bauer}},\
  }\href {https://doi.org/10.1103/PhysRevB.100.245121} {\bibfield  {journal}
  {\bibinfo  {journal} {Phys. Rev. B}\ }\textbf {\bibinfo {volume} {100}},\
  \bibinfo {pages} {245121} (\bibinfo {year} {2019})}\BibitemShut {NoStop}%
\bibitem [{\citenamefont {Meiburg}\ and\ \citenamefont
  {Bauer}(2022)}]{MeiburgBauer2022}%
  \BibitemOpen
  \bibfield  {author} {\bibinfo {author} {\bibfnamefont {A.}~\bibnamefont
  {Meiburg}}\ and\ \bibinfo {author} {\bibfnamefont {B.}~\bibnamefont
  {Bauer}},\ }\href {https://doi.org/10.1103/PhysRevResearch.4.023128}
  {\bibfield  {journal} {\bibinfo  {journal} {Phys. Rev. Research}\ }\textbf
  {\bibinfo {volume} {4}},\ \bibinfo {pages} {023128} (\bibinfo {year}
  {2022})}\BibitemShut {NoStop}%
\bibitem [{\citenamefont {White}(2017)}]{White2017Gausslets}%
  \BibitemOpen
  \bibfield  {author} {\bibinfo {author} {\bibfnamefont {S.~R.}\ \bibnamefont
  {White}},\ }\href {https://doi.org/10.1063/1.5007066} {\bibfield  {journal}
  {\bibinfo  {journal} {J. Chem. Phys.}\ }\textbf {\bibinfo {volume} {147}},\
  \bibinfo {pages} {244102} (\bibinfo {year} {2017})}\BibitemShut {NoStop}%
\bibitem [{\citenamefont {White}\ and\ \citenamefont
  {Stoudenmire}(2019)}]{WhiteStoudenmire2019}%
  \BibitemOpen
  \bibfield  {author} {\bibinfo {author} {\bibfnamefont {S.~R.}\ \bibnamefont
  {White}}\ and\ \bibinfo {author} {\bibfnamefont {E.~M.}\ \bibnamefont
  {Stoudenmire}},\ }\href {https://doi.org/10.1103/PhysRevB.99.081110}
  {\bibfield  {journal} {\bibinfo  {journal} {Phys. Rev. B}\ }\textbf {\bibinfo
  {volume} {99}},\ \bibinfo {pages} {081110} (\bibinfo {year}
  {2019})}\BibitemShut {NoStop}%
\bibitem [{\citenamefont {Sawaya}\ and\ \citenamefont
  {White}(2022)}]{sawayawhite}%
  \BibitemOpen
  \bibfield  {author} {\bibinfo {author} {\bibfnamefont {R.~C.}\ \bibnamefont
  {Sawaya}}\ and\ \bibinfo {author} {\bibfnamefont {S.~R.}\ \bibnamefont
  {White}},\ }\href {https://doi.org/10.1103/PhysRevB.105.045145} {\bibfield
  {journal} {\bibinfo  {journal} {Phys. Rev. B}\ }\textbf {\bibinfo {volume}
  {105}},\ \bibinfo {pages} {045145} (\bibinfo {year} {2022})}\BibitemShut
  {NoStop}%
\bibitem [{\citenamefont {Stoudenmire}\ and\ \citenamefont
  {White}(2017)}]{StoudenmireWhite2017}%
  \BibitemOpen
  \bibfield  {author} {\bibinfo {author} {\bibfnamefont {E.~M.}\ \bibnamefont
  {Stoudenmire}}\ and\ \bibinfo {author} {\bibfnamefont {S.~R.}\ \bibnamefont
  {White}},\ }\href {https://doi.org/10.1103/PhysRevLett.119.046401} {\bibfield
   {journal} {\bibinfo  {journal} {Phys. Rev. Lett.}\ }\textbf {\bibinfo
  {volume} {119}},\ \bibinfo {pages} {046401} (\bibinfo {year}
  {2017})}\BibitemShut {NoStop}%
\bibitem [{\citenamefont {Crosswhite}\ \emph {et~al.}(2008)\citenamefont
  {Crosswhite}, \citenamefont {Doherty},\ and\ \citenamefont
  {Vidal}}]{CrosswhiteDohertyVidal2008}%
  \BibitemOpen
  \bibfield  {author} {\bibinfo {author} {\bibfnamefont {G.~M.}\ \bibnamefont
  {Crosswhite}}, \bibinfo {author} {\bibfnamefont {A.~C.}\ \bibnamefont
  {Doherty}},\ and\ \bibinfo {author} {\bibfnamefont {G.}~\bibnamefont
  {Vidal}},\ }\href {https://doi.org/10.1103/PhysRevB.78.035116} {\bibfield
  {journal} {\bibinfo  {journal} {Phys. Rev. B}\ }\textbf {\bibinfo {volume}
  {78}},\ \bibinfo {pages} {035116} (\bibinfo {year} {2008})}\BibitemShut
  {NoStop}%
\bibitem [{\citenamefont {Lin}\ and\ \citenamefont {Tong}(2021)}]{LinTong2021}%
  \BibitemOpen
  \bibfield  {author} {\bibinfo {author} {\bibfnamefont {L.}~\bibnamefont
  {Lin}}\ and\ \bibinfo {author} {\bibfnamefont {Y.}~\bibnamefont {Tong}},\
  }\href {https://doi.org/10.1137/19M1287067} {\bibfield  {journal} {\bibinfo
  {journal} {SIAM J. Sci. Comput.}\ }\textbf {\bibinfo {volume} {43}},\
  \bibinfo {pages} {A164} (\bibinfo {year} {2021})}\BibitemShut {NoStop}%
\bibitem [{\citenamefont {Edelman}\ \emph {et~al.}(1998)\citenamefont
  {Edelman}, \citenamefont {Arias},\ and\ \citenamefont
  {Smith}}]{EdelmanAriasSmith1998}%
  \BibitemOpen
  \bibfield  {author} {\bibinfo {author} {\bibfnamefont {A.}~\bibnamefont
  {Edelman}}, \bibinfo {author} {\bibfnamefont {T.~A.}\ \bibnamefont {Arias}},\
  and\ \bibinfo {author} {\bibfnamefont {S.~T.}\ \bibnamefont {Smith}},\ }\href
  {https://doi.org/10.1137/S0895479895290954} {\bibfield  {journal} {\bibinfo
  {journal} {SIAM J. Matrix Anal. Appl.}\ }\textbf {\bibinfo {volume} {20}},\
  \bibinfo {pages} {303} (\bibinfo {year} {1998})}\BibitemShut {NoStop}%
\bibitem [{\citenamefont {White}(2026)}]{HFDMRGSoftware}%
  \BibitemOpen
  \bibfield  {author} {\bibinfo {author} {\bibfnamefont {S.~R.}\ \bibnamefont
  {White}},\ }\href
  {https://github.com/srwhite59/HFDMRG.jl/releases/tag/v0.1.0} {\bibinfo
  {title} {{HFDMRG.jl}}} (\bibinfo {year} {2026}),\ \bibinfo {note} {{Julia}
  software, version 0.1.0}\BibitemShut {NoStop}%
\end{thebibliography}%

\end{document}